\documentclass{ptephy_v1}

\newcommand{\beq}{\begin{eqnarray}}
\newcommand{\eeq}{\end{eqnarray}}
\def\tr{\mathop{\mathrm{tr}}\nolimits}

\begin{document} 
\title{Thermodynamics for pure SU($2$) gauge theory using gradient flow}

\author[1]{T.~Hirakida}
\affil[1]{Department of Physics, Graduate School of Sciences, Kyushu University, Fukuoka 819-0395, Japan
\email{hirakida@email.phys.kyushu-u.ac.jp}
}
\author[2,3]{E.~Itou}
\affil[2]{Research Center for Nuclear Physics (RCNP), Osaka University, Osaka, 567-0047, Japan}
\affil[3]{Department of Mathematics and Physics, Kochi University, 2-5-1 Akebono-cho, Kochi 780-8520, Japan
\email{itou@yukawa.kyoto-u.ac.jp}
}
\author[4]{H.~Kouno}
\affil[4]{Department of Physics, Saga University, Saga 840-8502, Japan
\email{kounoh@cc.saga-u.ac.jp}
}
\begin{abstract}
We study the equation of state of pure SU($2$) gauge theory using Monte Carlo simulations.
The scale-setting of lattice parameters has been carried by using the gradient flow.
We propose a reference scale $t_0$ for the SU($2$) gauge theory satisfying $t^2 \langle E \rangle|_{t=t_0} =0.1$,
which is fixed by a natural scaling-down of the standard $t_0$-scale for the SU($3$) case based on perturbative analyses.
We also show
the thermodynamic quantities as a function of $T/T_c$, which are derived by the energy-momentum tensor using the small flow-time expansion of the gradient flow.
\end{abstract}

\maketitle

\section{Introduction and motivation}
\label{sec:intro}

Thermodynamic quantities (energy, pressure, thermal entropy, {\it etc.}) and transport coefficients (heat capacity, shear and bulk viscosities, {\it etc.}) play an important role to understand features of Quantum ChromoDynamics (QCD).
In lattice {\it ab initio} calculations, the thermodynamic quantities~\cite{Boyd:1996bx, Borsanyi:2012ve} for the quenched QCD have been precisely obtained.
These precise thermodynamic data are well explained by the ideal gluon gas and the Hard-Thermal-Loop (HTL) model \cite{Andersen:2010ct} in the high temperature regime ($5T_c \lesssim T$) and by the massive free glueball model in $T \lesssim T_c$, where $T_c$ denotes the critical temperature of the confined/deconfined phase transition. 
The consistency between Lattice QCD and the model studies gives us an intuitive picture of QCD state in these two limits. 

As for the thermodynamic quantities, recently, Suzuki proposed a new
method to calculate the energy-momentum tensor (EMT) using the
gradient flow~\cite{Suzuki:2013gza}.
In the quenched QCD, the method works well for calculating the
pressure and the energy density,  and the statistical errors of
the thermodynamic quantities reduce in the gradient flow method~\cite{Asakawa:2013laa}.
Furthermore, the method has been also extended to the full QCD~\cite{Makino:2014taa} and the numerical simulations have been successfully performed~\cite{Itou:2015gxx,Taniguchi:2016tjc,Taniguchi:2016ofw}.

Other important quantities are the transport coefficients.
In the intermediate temperature, experimental data show the small shear viscosity-to-thermal entropy ratio ($\eta/s$), which is consistent with the most perfect-liquid property rather than the gas~\cite{Andronic:2014zha}. 
A theoretical large-$N_c$ analysis based on AdS/CFT correspondence gives the lower bound for $\eta/s$~\cite{Son:2007vk}, while the $1/N_c$ correction terms to $\eta/s$ have not yet been determined even for its sign in the finite $N_c$~\cite{Kats:2007mq}.
Although the determination of transport coefficients using the lattice calculations has been developing~\cite{Nakamura:2004sy, Meyer:2007ic,Astrakhantsev:2018oue, Pasztor:2018yae}, it is still a challenging subject.
The measurement of the correlation function of the energy-momentum tensor (EMT) is the first step to obtain the viscosities,
 and there are at least three difficulties: {\it (i)} the small signal-to-noise ratio of the correlator of EMT, {\it (ii)} the definition of the ``correct'' renormalization of EMT as a conserved quantity on the lattice, {\it (iii)} solving an inverse-problem to obtain the spectral function from the correlation function.
In fact, the recent study~\cite{Pasztor:2018yae} has used more than $6$-million configurations to obtain the shear viscosity for one set of lattice parameter in the pure SU($3$) gauge theory. 

Based on these situations, we focus on the pure SU($2$) gauge theory, which is a good testing ground for the SU($3$) gauge theory since the numerical cost is lower than the one for the SU($3$) because of the smallness of the matrix size, nevertheless it has almost the same properties as the SU($3$) gauge theory.
In addition, the study of SU($2$) gauge theory will provide the larger signal of the correction term of $1/N_c$ to $\eta/s$ because of the smaller $N_c$.

Several works have obtained the thermodynamic quantities for the SU($2$) gauge theory using the integral method~\cite{Engels:1990vr, Engels:1994xj, Caselle:2015tza,Giudice:2017dor} on the lattice.
A recent work, which mainly focuses on $T < T_c$, shows the consistency with the massive free glueball model~\cite{Caselle:2015tza}.
Another one~\cite{Giudice:2017dor}, which utilizes the improved gauge action, presents the thermodynamic quantities without taking continuum limit.
\begin{figure}[h]
 \centering
 \includegraphics[width=0.7\textwidth,clip]{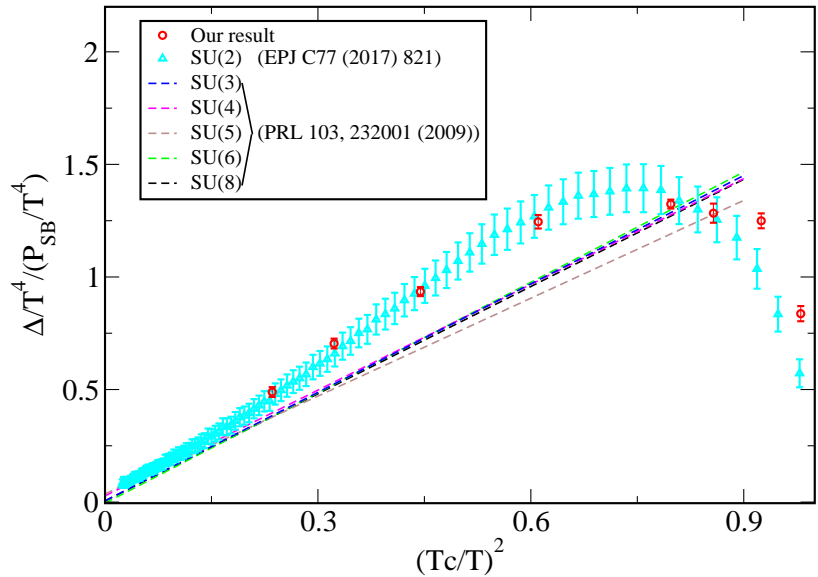}
 \caption{
The trace anomaly ($\Delta/T^4$) normalized by the pressure in the Stefan-Boltzmann (SB) limit ($P_{SB}/T^4$) as a function of $(T_c/T)^2$.
 The circle (red) and the triangle (cyan) symbols, which are given in this work and in Ref. \cite{Giudice:2017dor}, denote the results for the SU($2$) gauge theory, respectively.
 The dashed lines represent the interpolating functions of the results for several SU($N_c\ge 3$) gauge theories, which are given by Eq. (16) in Ref. \cite{Panero:2009tv}. The data of the SU($N_c \ge 3$) cases in the range $(T_c/T)^2 \le 0.9$ are well-fitted by the interpolating functions.
 }
 \label{fig:anomaly-vs-Panero}
\end{figure}
It is reported that, in intermediate temperature ($T_c \lesssim T \lesssim 5T_c$), the scaling law of the trace anomaly of the SU($2$) gauge theory has the different behavior from the SU($N_c$) gauge theories with $N_c \ge 3$ as shown in Fig.~\ref{fig:anomaly-vs-Panero}, where the $N_c$-dependence of the trace anomaly ($\Delta/T^4$) normalized by the pressure in the Stefan-Boltzmann (SB) limit ($P_{SB}/T^4 = \pi^2 (N_c^2 -1 )/45$) for the SU($N_c$) theories is plotted.
The results for the SU($2$) gauge theories present as the circle (red) symbols, which we precisely determine in this work, and the triangle (cyan) symbols obtained in Ref.~\cite{Giudice:2017dor}.
The dashed lines present the interpolating functions of the results for the SU($N_c$) gauge theories with $3 \le N_c \le 8$ given in Ref.~\cite{Panero:2009tv}.
In this plot, the horizontal axis denotes $(T_c/T)^2$.
The leading term of the trace anomaly is expected to be approximately proportional to $(T_c/T)^2$ in high temperature regime, and in fact, the results of the SU($N_c \ge 3$) theory exhibit the linear behavior in $(T_c/T)^2 \le 0.9$. 
(To say more exactly, trace anomaly is proportional to $g^4T^4$, where $g$ is the renormalized coupling constant, in the extremely high temperature limit $(T_c/T)^2\sim 0$~\cite{Kapusta:1979fh,Kapusta:finite,Romatschke:2009ng}.)
On the other hand, the data for the SU($2$) theory have a gently curved behavior in $ 0.6 \lesssim (T_c/T)^2 \lesssim 1.0$.
In this work, we show its scaling property very precisely after taking the continuum extrapolation.
We consider that the difference of the order of the phase transition between SU($2$) and SU($N_c \ge 3$) theories gives this dissimilar scaling behavior and it is a promising signal to find a large $N_c$-dependence for the thermodynamic quantities and transportation coefficients near $T_c$ through the study on the SU($2$) gauge theory.

In this work, we precisely determine the scale-setting function using $t_0$ reference-scale, which is available for $2.420 \le \beta \le 2.850$.
The function gives a key relationship  to determine the size of lattice spacing and the temperature.
Using the lattice parameters given by the scale-setting function, the thermodynamic quantities in $T_c \lesssim T \lesssim 2T_c$ are directly calculated from the EMT by the gradient flow method.
We carefully take the continuum limit of the lattice thermodynamic data and then compare the results to the other calculations; the numerical integral-method and the analytical HTL calculations.
Several independent methods are proposed to determine the equation of state by using the numerical simulations; the integral method~\cite{Engels:1990vr}, the moving-frame method~\cite{Giusti:2016iqr}, the gradient flow method~\cite{Asakawa:2013laa}, and the non-equilibrium methods~\cite{Caselle:2018kap}, and it can extend these methods to the QCD including dynamical fermions~\cite{Heller:1984eq,DallaBrida:2017sxr, Itou:2015gxx, Borsanyi:2010cj}. 
Among them, the gradient flow method has two advantages; the statistical uncertainties are smaller and the wave-function renormalization of the EMT is not necessary for pure gauge theories. 
This is a promising approach to overcome difficulties {\it (i)} and {\it (ii)}.

The structure of the paper is following:
In \S.~\ref{sec:definition}, we briefly review the gradient flow equation.
We propose a reference scale for the scale setting using the flowed action density  in the SU($2$) gauge theory, which is obtained by a natural scale-down of the standard $t_0$ scale for the SU($3$) theory.
Furthermore, we give a brief review of the calculation for the EMT using the flowed field variable.
In \S.~\ref{sec:simulation-setup}, the simulation setup for the configuration generation and how to solve the gradient flow equation are explained.
In \S.~\ref{sec:scale-setting}, we show our results for zero-temperature simulations. We obtain the scale-setting equation, which gives the relationship between the lattice bare coupling constant and the lattice spacing.  
The ratios between our reference-scale and the other quantities are calculated.
In \S.~\ref{sec:finite-temp}, the results for the finite-temperature simulations are shown.
We obtain the thermodynamic quantities as a function of temperature and compare the results with the other analyses.
Section~\ref{sec:summary} gives the summary and future directions of this work.

\section{Definition of the observables}\label{sec:definition}
\subsection{Reference scale for the scale-setting equation using the gradient flow}\label{sec:gradient-flow}
The Yang-Mills gradient flow equation, which is a key equation in this work, is defined by
\beq
\frac{\partial B_\mu}{\partial t} = D_\nu G_{\nu \mu},~~~B_\mu(t,x)|_{t=0} = A_\mu(x), \label{eq:Wilson-flow}
\eeq
where $t$ denotes a fictitious time, which is called ``flow-time'', and $A_\mu$ and $B_\mu$ are quantum and deformed gauge fields by the flow in the SU($N_c$) gauge theory, respectively. 
Note that the flow-time has mass dimension $-2$. 
The operator $G_{\mu\nu}$ is the field strength consisting of $B_\mu$. Thus the right hand side in Eq.~(\ref{eq:Wilson-flow}) shows the same form as the equation of motion for the deformed gauge field $B_\mu$. 
The solution of the equation parametrized by the flow-time defines a transformation of the gauge field
toward the stationary points of the gauge action.
At the time~$t$, the high frequency mode, whose momentum is larger than $1/\sqrt{t}$, is suppressed (see Eq.~($2.18$) in Ref.~\cite{Luscher:2010iy}). 
The deformed field can be considered to the renormalized field by the nonperturbative transformation and the flow-time can be identified as a typical energy scale of the renormalization.

Now, we introduce a length scale of the order of $ \sqrt{t} $ to the local operators through the flow equation.
This scale gives the renormalization condition to the local operators
 in spite of the fact that the operator itself has no characteristic length.
Moreover, the flows do not separate the modes of the gauge fields in contrast to
 the case of the ordinary cutoff scale, and the gauge invariance is preserved.
The one-loop computation~\cite{Luscher:2010iy} shows that the energy density $ E = (1/4) G^a_{\mu\nu}G^a_{\mu\nu}$ can be
renormalized at the momentum scale of $ 1/ \sqrt{8t}$ and can be represented by the renormalized coupling constant;
\beq
\langle E(t) \rangle &=& \frac{3 (N_c^2-1)}{32 \pi t^2} \alpha_s(\mu^2_R) \left[ 1+ k_1 \alpha_s (\mu^2_R) + O(\alpha^2))  \right],\label{eq:pert-t2E}\\
k_1&=& \frac{1}{4\pi} \left[ N_c \left( \frac{11}{3} \gamma_E +\frac{52}{9} -3 \ln 3 \right) -N_f \left( \frac{2}{3}\gamma_E + \frac{4}{9} -\frac{4}{3} \ln 2 \right) \right].
\eeq
Here, $\gamma_E$ denotes the Euler constant and $\alpha_s$ is the running coupling constant, $\alpha_s = g^2/(4\pi)$, at the scale $\mu_R$ with the renormalized coupling constant $g$ in the $\overline{\rm{MS}} $ scheme.
The next-next-to-leading order calculation of the $E$ is given in Ref.~\cite{Harlander:2016vzb}.

The $\beta$-function for the SU($N_c$) gauge theory is given by the following form;
\beq
\mu_R^2 \frac{d \alpha_s}{d \mu_R^2}= - ( b_0 \alpha_s^2 + b_1 \alpha_s^3 + b_2 \alpha_s^4 + b_3 \alpha_s^5 +O(\alpha_s^6) ),
\eeq
where  the coefficients $b_0$ -- $b_3$ are calculated in Ref.~\cite{vanRitbergen:1997va};
\beq
b_0&=& \frac{1}{4 \pi}\frac{11N_c}{3},~~b_1=\frac{1}{(4\pi)^2} \frac{34}{3} N_c^2,~~b_2=\frac{1}{(4\pi)^3} \frac{2857}{54}N_c^3,\nonumber\\
b_3&=&\frac{1}{(4\pi)^4}\left[ \left(\frac{150653}{486} -\frac{44}{9} \zeta_3 \right) N_c^4 +\frac{N_c^2 (N_c^2+36)}{24} \left( -\frac{80}{9} +\frac{704}{3} \zeta_3  \right) \right]. \nonumber\\ \label{eq:coeff-b}
\eeq
Here, $\zeta$ is the Riemann zeta-function and $\zeta_3=1.2020569\cdots$.
It is well-known that the $\beta$-function does not depend on the renormalization scheme up to two-loop order.

The solution of the $\beta$-function gives the running coupling constant, which is explicitly written by
\beq
\alpha_s(\mu_R^2) &\simeq& \frac{1}{b_0 \tilde{t}} \left( 1- \frac{b_1}{b_0^2} \frac{\ln \tilde{t}}{\tilde{t}} + \frac{b_1^2 (\ln^2 \tilde{t} -\ln \tilde{t} -1) +b_0 b_2}{b_0^4 \tilde{t}^2} \right. \nonumber\\
&& \left.- \frac{b_1^3 ( \ln^3 \tilde{t} -\frac{5}{2} \ln^2 \tilde{t} -2 \ln \tilde{t} +\frac{1}{2} ) +3b_0 b_1 b_2 \ln \tilde{t} -\frac{1}{2} b_0^2 b_3 }{ b_0^6 \tilde{t}^3}   \right) ,
\label{eq:alpha-s}
\eeq
where we put $\tilde{t} \equiv \ln \frac{\mu_R^2}{\Lambda^2}$, and $\mu_R^2 =1/(8t)$ at the gradient flow-time $t$.
Once we fix the $\Lambda_{\overline{\mathrm{MS}}}$, we can estimate the expectation value of $E(t)$ as a function of flow-time. 
Note that $E(t)$ in Eq.~(\ref{eq:pert-t2E}) takes a finite value in non-zero flow-time even in the continuum theory. 

For the SU($3$) gauge theory, a new reference scale, namely $t_0$-scale, has been proposed in Ref.~\cite{Luscher:2010iy},
\beq
t^2 \langle E(t) \rangle |_{t=t_0} = 0.3~~~~~\mbox{for SU($3$),}
\eeq 
where $t^2 E(t)$ is a dimensionless quantity.
One advantage of the usage of this reference scale is less statistical uncertainties since the operator $E$ is a local operator. 
In general, we can take any reference value of $A (\equiv t^2 \langle E(t) \rangle$).
The discretization error becomes larger for smaller $A$. The statistical error and the numerical cost grow up for the larger $A$.
In $A=0.3$ for the SU(3) gauge theory, the $t_0$-scale gives almost the same scale as the Sommer scale ($r_0$)~\cite{Luscher:2010iy}, and it is a suitable reference scale to investigate the thermal phase transition in the SU($3$) gauge theory.

Similar scales for the SU($2$) gauge theory with the definition $t^2 \langle E \rangle =0.2$ and $0.3$~\cite{Giudice:2017dor} and the others~\cite{Berg:2016wfw} have been discussed.
In this work, we would like to introduce the $t_0$-scale in the SU($2$) gauge theory as
\beq
t^2 \langle E(t) \rangle |_{t=t_0} = 0.1 ~~~~~\mbox{for SU($2$).}\label{eq:def-t0}
\eeq 
The perturbative analysis (Eq.~(\ref{eq:pert-t2E})) shows that the coefficient of $\alpha_s$ in the leading order is proportional to the number of gauge bosons.
The reference value in Eq.~(\ref{eq:def-t0}) is chosen as an approximate scaling-down of the standard $t_0$-scale in the SU($3$) theory.
We will show that our $t_0$-scale is almost the same scale with the Necco-Sommer scale ($r_c$) and is good for the study of thermodynamics in the temperature region of $T_c\lesssim T \lesssim 2T_c$.

\subsection{Thermodynamic quantities from the energy-momentum tensor}
In this work, we will obtain the thermodynamic quantities from the energy-momentum tensor (EMT).
In general, measurements of the EMT using the lattice numerical simulation are essentially difficult, since 
the lattice regularization manifestly breaks the general covariance, while EMT is a generator of the  corresponding invariance~\cite{Caracciolo:1990emt, Giusti:2015daa}.
Here, we calculate the EMT by using a new technique~\cite{Asakawa:2013laa} based on the small flow-time expansion  of the Yang-Mills gradient flow~\cite{Luscher:2010iy, Suzuki:2013gza}.

The key relation is given in Ref.~\cite{Suzuki:2013gza} for the quenched QCD based on the small flow-time expansion.
\beq
   T_{\mu\nu}^R(x)
   =\lim_{t\to0}\left\{\frac{1}{\alpha_U(t)}U_{\mu\nu}(t,x)
   +\frac{\delta_{\mu\nu}}{4\alpha_E(t)}
   \left[E(t,x)-\left\langle E(t,x)\right\rangle_0 \right]\right\},
\label{eq:def-EMT}
\eeq
where $\langle E \rangle_0$ is the expectation value of the trace-part of the EMT.
Here, we define the ``correctly-renormalized conserved EMT'' ($T_{\mu \nu}^R (x)$) by the subtraction of vacuum expectation values (v.e.v.) of the EMT.

Let us briefly review how to obtain the relation.
There are two gauge-invariant local products of dimension-$4$, $U_{\mu \nu}(t,x)$ and $E(t,x)$, in finite flow-time. 
These composite operators are UV finite in the positive flow-time ($t>0$) and are manifestly given by
\beq
U_{\mu \nu}(t,x)&=&G^a_{\mu \rho} G^a_{\nu \rho}(t,x) -\frac{1}{4}\delta_{\mu \nu} G^a_{\rho \sigma} G^a_{\rho \sigma}(t,x), \nonumber\\
E(t,x)&=&\frac{1}{4} G^a_{\mu \nu}G^a_{\mu \nu}(t,x).\label{eq:def-U-E}
\eeq
As an advantage of the usage
of the gradient flow, it is not necessary to calculate the wave function
renormalization of operators thanks to their UV finiteness in pure gauge 
theories~\cite{Luscher:2011bx}.
They can expand the dimension-$4$ and gauge covariant operators in terms of small flow-time in continuum theory;
\beq
U_{\mu \nu} (t,x)&=& \alpha_U (t) \left[ \{ T_{\mu \nu}^R (x) \} -\frac{1}{4} \delta_{\mu \nu} \{ T_{\rho \rho}^R (x) \} \right] + \mathcal{O}(t),\nonumber\\
E(t,x)&=& \langle E(t,x) \rangle_0 + \alpha_E (t) \{ T_{\rho \rho}^R (x) \} + \mathcal{O}(t).\label{eq:E-expand-t} 
\eeq
The EMT ($T_{\mu \nu}^R$) is a conserved quantity, so that it must be a scheme-independent.
Thus, the coefficients ($\alpha_U (t)$ and $\alpha_E(t)$) should be also scheme-independent.
However, it is hard to determine the coefficients nonperturbatively, so that we practically utilize the calculated values in the one-loop order in Ref.~\cite{Suzuki:2013gza},
\beq
\alpha_U(t)&=&4 \pi \alpha_s \{ 1+ 2 b_0 \bar{s}_1 \alpha_s +O(\alpha_s^2) \},  \\
\alpha_E(t)&=&\frac{2 \pi}{b_0} \{ 1+ 2 b_0 \bar{s}_2 \alpha_s +O(\alpha_s^2) \},
\eeq
where $\alpha_s$ is the renormalized coupling constant in the $\overline{\rm{MS}} $ scheme (Eq.~(\ref{eq:alpha-s})) at the scale $\mu_R=1/\sqrt{8t}$. 
The coefficients $\bar{s}_1$ and $\bar{s}_2$ depend on the renormalization scheme, which is the same as the scheme for $\alpha_s$.
In the $\overline{\rm{MS}} $ scheme,
\beq
\bar{s}_1&=&\frac{7}{22 } +\frac{\gamma_E}{2} -\ln 2 \simeq -0.08635752993, \\
\bar{s}_2&= &\frac{21}{44}-\frac{b_1}{2b_0^2} \simeq 0.0557812397.
\eeq
Although the coefficients are derived perturbatively, we assume that the relation is available in the nonperturbative regime.
Then, we can obtain the renormalized EMT from Eq.~(\ref{eq:def-EMT}), if we can calculate the values of $U_{\mu \nu}$ and $E$ in a nonperturbative method, for instance, the lattice simulation.
In practice, the operators $U_{\mu \nu}$ and $E$ can be calculated by the clover-leaf operators on the lattice.

In the finite-temperature system, the following components of the EMT correspond to a combination of the energy density ($\varepsilon$) and the pressure ($P$) at the temperature $T$,
\beq
\varepsilon -3P &=& - \sum_{\mu=1}^{4} T_{\mu \mu}, \\
 sT &=&  \varepsilon +P =  T_{11}-T_{44} .
\eeq
Here, $s$ denotes the thermal entropy density, and the EMT at the temperature $T$ is given by
\beq
   T_{\mu\nu}(x)
   &=&\lim_{t\to0}\left\{\left.\frac{1}{\alpha_U(t)}U_{\mu\nu}(t,x)\right|_T \right. 
   \left. +\frac{\delta_{\mu\nu}}{4\alpha_E(t)}
  \left[ \left[E(t,x)\right]_{T}  - \langle E(t,x)\rangle_{T=0} \right] \right\}.
\eeq

\section{Simulations setup}\label{sec:simulation-setup}
\subsection{Configuration generation}
We utilize the standard Wilson-Plaquette gauge action defined on a 
four-dimensional Euclidean lattice,
\beq
S_W=  \frac{2N_c}{g_0^2} \sum_{x, \mu > \nu}  \left(1- \frac{1}{N_c}\tr P_{\mu \nu} (x)\right).\label{eq:Wilson-action}
\eeq
Here, $g_0$ and $P_{\mu \nu}$ denote the lattice bare coupling constant and the plaquette,
\beq
P_{\mu \nu}(x)=U_{\mu} (x) U_{\nu} (x+a\hat{\mu}) U^\dag_{\mu}(x+a\hat{\nu}) U^\dag_\nu (x),
\eeq
respectively.
Here $U_{\mu}(x)$ represents the link-variable from a site $x$ to its neighbor in $\mu$-direction and takes the value with the SU($N_c$) group elements.
Hereafter, we introduce the lattice parameter $\beta=2N_c/g_0^2$.
 
In this work, we adopt periodic boundary conditions for all directions.
Gauge configurations are generated by using the pseudo-heatbath algorithm with 
over-relaxation.
We use the word ``a sweep'' to refer to the combination of one pseudo-heatbath 
update step followed by these multiple over-relaxation steps. 
The mixed ratio of the combination is $1$:$20$ for zero-temperature configuration, and for finite temperature, the ratio is $1$:$N_\tau$.
In order to eliminate the influence of autocorrelation, we take large enough number of sweeps ($100$ sweeps) 
between measurements.
To see that, we observe the topology changing between each measured configuration (see Appendix~\ref{sec:topology}). 
The number of configuration is $100$--$600$ and the lattice extent is $N_s^4=32^4$ for the zero-temperature simulation.
For the finite-temperature simulation,  the number of configurations is $200$ on $N_s^3 \times N_\tau$ with the fixed aspect ratio $N_s/N_\tau=4$ for $N_\tau=6,8,10$, and $12$.

\subsection{Gradient flow equation on the lattice}
In \S.~\ref{sec:gradient-flow}, we briefly explained how the gradient flow works in the continuum theory. 
The method is also applicable to lattice formulations.
The flow equation on the lattice is given in Refs.~\cite{Luscher:2009eq,Luscher:2010iy},
\beq
\partial_t V_t(x,\mu)=-g_0^2 \{ \partial_{x,\mu}S_W(V_t) \} V_t(x,\mu), ~~~ V_t(x,\mu)\bigl|_{t=0}=U_{\mu}(x).
\eeq
Here, $ U_\mu(x) $ is the $ \mathrm{SU}(N_c) $ link variable and $V_t$ is the deformed one.
The $ S_W$ in the right hand side denotes the Wilson-Plaquette action.
The uniqueness and smoothness of the gradient flow on the lattice are guaranteed.
Furthermore, the expectation value of a local operator after the flow by the lattice simulation has a well-defined continuum limit.

The equation is an ordinary first-order differential equation so that we numerically integrate it using the Runge-Kutta (RK) method.
The third-order RK formula with the initial configuration $V_0=U_{\mu}(x)$ was given in Ref.~\cite{Luscher:2010iy}.
Each step of the integration is obtained as
\beq
W_0 &=& V_t, \nonumber \\
W_1 &=& \mathrm{exp} \{ \tfrac{1}{4} Z_0 \} W_0, \nonumber \\
W_2 &=& \mathrm{exp} \{ \tfrac{8}{9} Z_1- \tfrac{17}{36} Z_0 \} W_1, \nonumber \\
V_{t+\epsilon} &=& \mathrm{exp} \{ \tfrac{3}{4} Z_2 - \tfrac{8}{9}Z_1 +\tfrac{17}{36} Z_0 ) \} W_2,
\eeq
where
\beq
Z_i&=&\epsilon Z(W_i),  \ \ \  i=0,1,2,\\
Z(W_i) &\equiv& -g_0^2 \{ \partial_{x,\mu}S_W(W_i) \}.
\eeq
The error per step is of order $\epsilon^3$ in the third-order RK method.
 We take $\epsilon=0.01$.

\section{Scale setting}\label{sec:scale-setting}
\subsection{Lattice data of $t^2 \langle E (t) \rangle$ in perturbative regime }
Before carrying out the scale setting simulation, we compare the operator $E$ as a function of the flow-time between the perturbative result in the continuum theory and the numerical data on the lattice. 
To set the lattice parameter in the perturbative regime, we take $\beta=2.85$.

The operator $E= G_{\mu \nu}^a G_{\mu \nu}^a /4 $ can be constructed by the clover-leaf operator on the lattice.
Furthermore, there is the other simpler definition; 
\beq
E = 2 \sum_{(x \in P_x)} \Re \tr \left(1- \tilde{P}_{\mu \nu}^t (x)\right), \label{eq:plaq-E}
\eeq
where $\tilde{P}_{\mu \nu}^t (x)$ represents a flowed-plaquette and $P_x$ is the set of unoriented plaquette in which the site of lower-left corner is $x$.
In the continuum limit, the results should be independent of the definitions on the lattice.

It is worth to summarize several values for the SU($2$) gauge theory with the Wilson-Plaquette action on the lattice.

The running coupling constant in the Schr\"{o}dinger function (SF) scheme for the SU($2$) gauge theory has been investigated in Ref.~\cite{deDivitiis:1994yz}.
The nonperturbative $\beta$-function is estimated by the following polynomial
\beq
L \frac{\partial g_{SF}}{\partial L} &=& \frac{b_0}{4 \pi} g^3_{SF}+ \frac{b_1}{(4 \pi)^2} g^5_{SF} +\frac{b_2^{eff}}{(4 \pi)^3} g^7_{SF}, \label{eq:nonpert-beta}\\
b^{eff}_2& =& 0.35(12). \nonumber
\eeq
Here, $L$ denotes the scale in the finite volume scaling and $b_2^{eff}$ is determined by fitting the nonperturbative running coupling constant with the functional form given in Eq.~(\ref{eq:nonpert-beta}).
On the large volume, where $g^2_{SF} (L_0^{SF})=4.765$, we can calculate $L_0^{SF} \Lambda_{SF}$ by solving the $\beta$ function;
\beq
L_0^{SF} \Lambda_{SF} = 0.1804^{+0.0101}_{-0.0064}.
\eeq
On the tree-level SF action with $\beta=2.85$, the value of renormalized coupling $g^2_{SF} =4.765 $ is realized on $a/L_0^{SF}=0.0834(5)$.
Using the ratio of $\Lambda$ parameters between SF and $\overline{\mathrm{MS}}$ schemes,
${\Lambda_{SF}}/{\Lambda_{\overline{\mathrm{MS}}}} = 0.44567$~\cite{deDivitiis:1994yz},
we obtain the $\Lambda_{\overline{\mathrm{MS}}}$ in lattice units at $\beta=2.85$;
\beq
a(2.85) \Lambda_{\overline{\mathrm{MS}}}=0.0338^{+0.0021}_{-0.0014}. \label{eq:Lambda-MSbar}
\eeq
\begin{figure}[h]
\begin{center}
\includegraphics[width = 0.6\textwidth,clip]{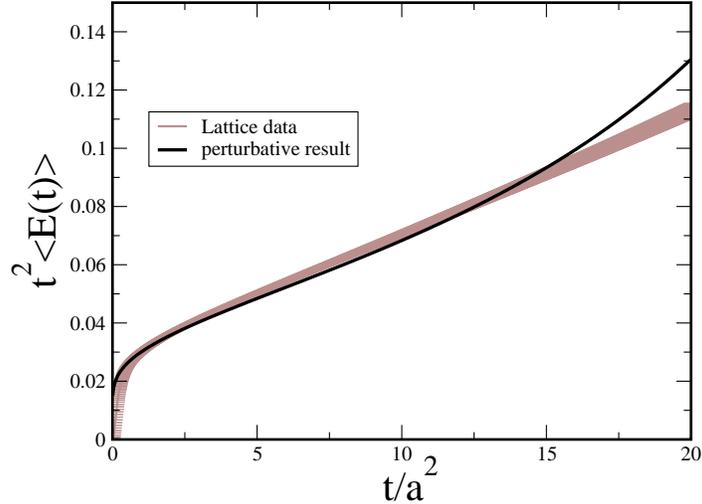}
\caption{Comparison $t^2 \langle E (t) \rangle$ data between the perturbative calculation (Eq.~(\ref{eq:pert-t2E})) and the lattice result at $\beta=2.85$.}
\label{fig:t2E-beta2.85}
\end{center}
\end{figure}

Figure~\ref{fig:t2E-beta2.85} shows $t^2 \langle E \rangle$ as a function of $t/a^2$, where the operator $E$ on lattice is calculated by the clover-leaf.
The thick (brown) bound denotes the lattice numerical result with $1$--$\sigma$ statistical error bar, and the black solid curve is the perturbative calculation using Eqs.~(\ref{eq:pert-t2E}) and~(\ref{eq:Lambda-MSbar}).
Here, the flow-time corresponds to the renormalization scale, so that the small flow-time regime must deal with the perturbative behavior.
We confirm that the lattice and the perturbative calculations are consistent with each other in the small flow-time regime, as expected.

\subsection{$t_0$-scale in the SU($2$) gauge theory}
\begin{figure}[h]
\begin{center}
\includegraphics[width = 0.6\textwidth,clip]{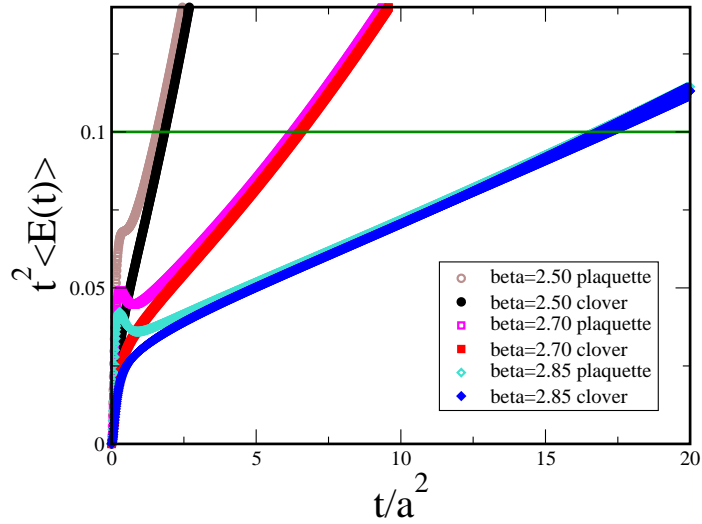}
\caption{Flow-time dependence of $t^2 \langle E (t) \rangle$ at $\beta=2.500, 2.700$ and $2.850$. Thinner-open symbols (brown, magenta, cyan) are obtained by the plaquette definition of the operator $E$, while thicker-filled ones (black, red, blue) are given by the clover definition. }
\label{fig:t2E-beta-deps}
\end{center}
\end{figure}

Now, we measure the $t^2 \langle E(t) \rangle$ in the range of $\beta=2.400$--$2.900$ on $32^4$ lattice.
Figure~\ref{fig:t2E-beta-deps} is the plot for the expectation values of $t^2 \langle E(t) \rangle$ at $\beta=2.500,2.700$ and $2.850$.
Here, we show two types of the data for each $\beta$; one is given by the plaquette definition (Eq.~(\ref{eq:plaq-E})) as the lattice operator $E$, while the other is obtained from the clover definition.
It is known that there is a linear-scaling region of $t^2\langle E(t) \rangle$ as a function of $t$ for the SU($3$) gauge theory.
We found that the data in the SU($2$) gauge theory also present the same property.

The values of $t_0$ in lattice units are summarized in Table~\ref{table:t0-beta}.
\begin{table}[h]
\begin{center}
\begin{tabular}{|c|c|c|}
\hline
 $\beta$  &  $t_0/a^2$ & \# of Conf.   \\
 \hline
2.400  & 0.9549(5) & 300\\    
2.420  & 1.083(2)   & 100 \\      
2.500  & 1.839(3) &   300\\      
2.600  & 3.522(10) & 300\\      
2.700  & 6.628(36) &  300\\      
2.800  & 11.96(12) & 300\\          
2.850  & 16.95(17) & 600\\ 
\hline
\end{tabular}
\caption{$t_0/a^2$ and the number of configurations for each $\beta$.
} \label{table:t0-beta}
\end{center}
\end{table}
At $\beta=2.400$, the value of $t_0$ in lattice units is smaller than unity.
To avoid a strong lattice artifact, we drop the data at $\beta=2.400$ in the following analysis.
On the other hand, to avoid a finite volume effect we concentrate on $\sqrt{8t/a^2} \le 32/2$ regime, since beyond the regime all link variables on the lattice are averaged under the periodic boundary condition.
We also calculate the data at $\beta=2.900$, but it does not grow up larger than the reference value in the regime so that we do not include the data at $\beta=2.900$ in our analysis.

The data of $\ln(t_0/a^2)$ are well interpolated using a quadratic function of $\beta$.
The parametrization of $\ln(t_0/a^2)$ is determined by the best fit-function;
\beq
\ln(t_0/a^2)=1.258+ 6.409 (\beta-2.600) -0.7411 (\beta-2.600)^2,\label{eq:t0-beta}
\eeq
for $2.420 \le \beta \le 2.850$.
This scale-setting equation gives the relation between the lattice spacing ($a$) and  the temperature for a given $\beta$.

\begin{figure}[h]
\begin{center}
\includegraphics[width=0.6\textwidth,clip]{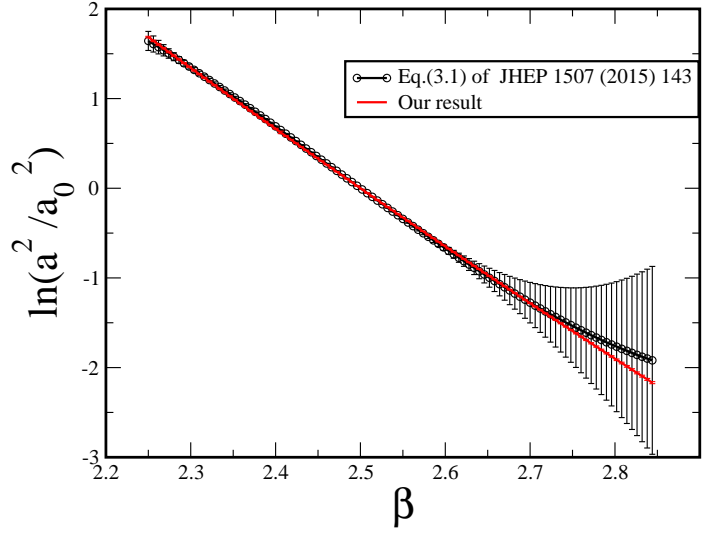}
\caption{ Ratio of the lattice spacing ($\ln(a^2/a_0^2)$) as a function of $\beta$. Here, $a_0$ is the lattice spacing at $\beta=2.50$. The black data are calculated by the scale-setting function Eq.(3.1) in Ref.~\cite{Caselle:2015tza}, and the red data are given by Eq.~(\ref{eq:t0-beta}). The error bars of our result are the same with the symbol size. }
\label{fig:ratio-latt-space}
\end{center}
\end{figure}
Figure~\ref{fig:ratio-latt-space} shows the ratio of the lattice spacing $\ln (a^2/a_0^2)$, where we take the reference lattice spacing $a_0$ with the value at $\beta=2.50$.
As a comparison, the previous data, which are obtained from the scale-setting equation in Ref.~\cite{Caselle:2015tza}, are also shown;
\beq
\ln (\sigma a^2)=-2.68-6.82(\beta-2.40)-1.90(\beta-2.40)^2+9.96(\beta-2.40)^3, ~~~~
\eeq
for  $2.27 \le \beta \le 2.60$.
Here, we estimate the error bar by using the covariance matrix of the chi-square fit.
There is a precise agreement between the scalings given by two scale-setting functions within $1$--$\sigma$ error bar in $2.42 \le \beta \le 2.60$, where both functions are available.
Our data cover the higher $\beta$ regime, which enables us to investigate the physics in $T_c \lesssim T$.

We also do a similar calculation for the other reference scales.
The results are summarized in Appendix~\ref{sec:other_t0_scale}.

\subsection{Relationship between $t_0$ and the other reference scales}
\begin{figure}[h]
\begin{center}
\includegraphics[width=0.9\textwidth,clip]{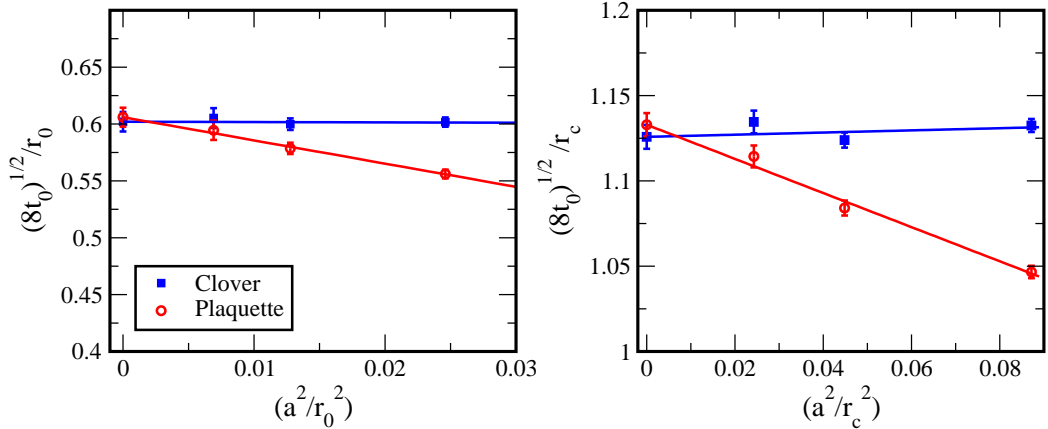}
\caption{ Left (right) panel shows the continuum extrapolation for the ratio between $\sqrt{8t_0}$ and $r_0$ ($r_c$). Filled-square (blue) data are obtained from the clover representation in the calculation of $E$, while the open-circle (red) data are calculated using the plaquette representation in both panels. }
\label{fig:comp-t0-r0}
\end{center}
\end{figure}

The relationship between the following typical scales in the theory is useful to understand the dynamics.
Left (right) panel in Fig.~\ref{fig:comp-t0-r0} shows the continuum extrapolation of the ratio between $\sqrt{8t_0}$ and  Sommer scale ($r_0$)~\cite{Sommer:1993ce} (Necco-Sommer scale ($r_c$)~\cite{Necco:2001xg}). 
Here, $r_0$ and $r_c$ are defined via the static quark-antiquark force;
$r^2 F(r)|_{r=r_0}=1.65$ and $r^2 F(r)_{r=r_c}=0.65$, respectively.
We estimate these values in lattice units using the data of $a^2 F(r)$ at $\beta=2.50, 2.60,$ and $2.70$ shown in Table~1 of Ref.~\cite{Sommer:1993ce}.

The filled-square (blue) and open-circle (red) symbols in each panel are obtained as a function of $t_0/a^2$ by using the clover and plaquette definitions for the operator $E$, respectively.
It is shown that the ratio of $t_0$ to $r_0$ or $r_c$ in the continuum limit takes the universal value in spite of the lattice definition of the operator.
The gentle slope of the data obtained by the clover-definition indicates a small discretization error. 
In fact, the size of order $a$ effects depends on the combination of the lattice action in the configuration generation, the lattice action in the gradient flow, and the lattice definition of the operators.
The discretization effects for these choices in the tree-level are systematically discussed in Refs.~\cite{Fodor:2014cpa, Ramos:2015baa}.

The values in the continuum limit with the clover-definition of the operator $E$ are given by
\beq
\frac{\sqrt{8t_0}}{r_0} = 0.6020(86)(40),~~~~~~~~~~~~~\frac{\sqrt{8t_0}}{r_c}=1.126(7)(7),\label{eq:data-t0-r0}
\eeq
where the first bracket denotes the statistical $1$--$\sigma$ error, and the second one shows the systematic error estimated by the deviation from the value calculated in the plaquette representation.
The discretization effect is well under control since the systematic uncertainty is smaller than the statistical error.
Although in the SU($3$) gauge theory, the $t_0$-scale (actually $\sqrt{8t_0}$) is roughly a similar-scale with the Sommer scale (See Fig.~3 in Ref.~\cite{Luscher:2010iy}), our $t_0$-scale in the SU($2$) gauge theory is closer to the Necco-Sommer scale ($r_c$).
Now, it is easy to obtain the scale in ``physical'' units. If we take $r_0=0.50$ [fm], then $\sqrt{8t_0}=0.3010(43)(20)$ [fm].

\begin{figure}[h]
\begin{center}
 \includegraphics[width=0.6\textwidth,clip]{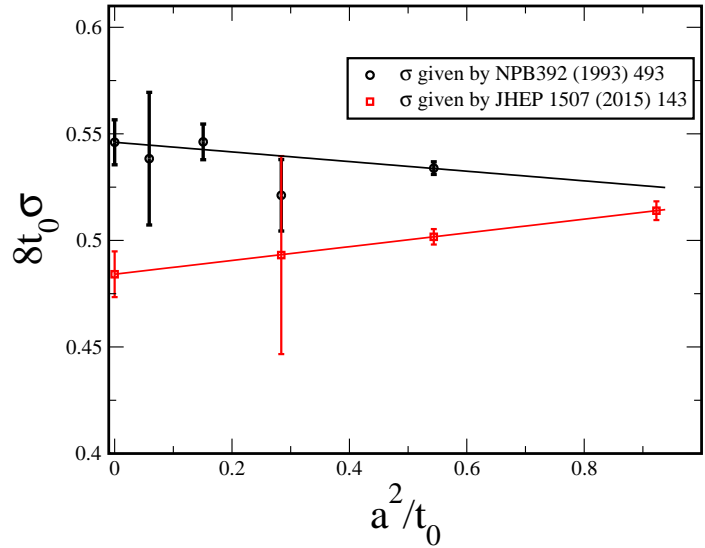}
\caption{ Continuum extrapolation of $8 t_0 \sigma$. Horizontal axis denotes $(a^2/t_0)$. The data, $\sigma a^2$, of the circle (black) symbol are listed in Table~4 in Ref.~\cite{Fingberg:1992ju}, while the ones of the square (red) symbol are given in Table~1 of Ref.~\cite{Caselle:2015tza} . }
\label{fig:t0sigma}
\end{center}
\end{figure}
As a consistency check, we consider $\sigma r_0^2$ via our $t_0$ scale.
We utilize the data of $a^2 \sigma$, where $\sigma$ denotes the string tension, given in Table~1 in Ref.~\cite{Caselle:2015tza} and Table~$4$ in Ref.~\cite{Fingberg:1992ju}.
Figure~\ref{fig:t0sigma} shows the continuum extrapolation of $8t_0\sigma$.
The circle (black) data are obtained by using our $8t_0/a^2$ and $\sigma a^2$ in Ref.~\cite{Fingberg:1992ju} at $\beta=2.50,2.60, 2.70$, and $2.85$, while the square (red) symbols are calculated by using the same $8t_0/a^2$ and $\sigma a^2$ in Ref.~\cite{Caselle:2015tza} at $\beta=2.42,2.50,$ and $2.60$.
The former data give $8t_0\sigma=0.546(11)$ and the latter ones show 
\beq
8t_0 \sigma=0.484(11).\label{eq:8t0sigma}
\eeq
It shows the $6$--$\sigma$ consistency, and the difference mainly comes from the discrepancy of the data of $a^2 \sigma$ at $\beta=2.50$ between Refs.~\cite{Caselle:2015tza} and ~\cite{Fingberg:1992ju}.
Using these $8t_0 \sigma$ and Eq.(\ref{eq:data-t0-r0}), we obtain $\sigma r_0^2=1.51(8)$ by the former data, while $\sigma r_0^2= 1.34(7)$ by the latter data.
In Ref.~\cite{Sommer:1993ce}, the value is independently obtained as $\sigma r_0^2=1.39(50)$ via the lattice-size as a reference scale, so that it is consistent with our value within $1$--$\sigma$ error bar, which is obtained via the $t_0$-scale.

Finally, using $T_c/\sqrt{\sigma}=0.69(2)$, $T_c/\Lambda_{{\overline{\mathrm{MS}}}}=1.23(11)$, $\sqrt{\sigma}/\Lambda_{{\overline{\mathrm{MS}}}}=1.79(12)$~\cite{Fingberg:1992ju}, and Eq.~(\ref{eq:8t0sigma}), we can calculate the relationship between our $t_0$ scale and the critical temperature ($T_c$) and the QCD scale in $\overline{\mathrm{MS}}$ scheme ($\Lambda_{\overline{\mathrm{MS}}}$) in the SU($2$) gauge theory;
\beq
\sqrt{8t_0}T_c=0.480(20),~~~~~\sqrt{8t_0} \Lambda_{{\overline{\mathrm{MS}}}}=0.389^{+0.032}_{-0.029}.
\eeq
Our $t_0$-scale is good for the study of thermodynamics in the temperature region of $T_c\lesssim T \lesssim 2T_c$, since $T_c<1/r_0<1/\sqrt{8t_0}(\sim 2T_c)<1/r_c$.

\section{Finite temperature}\label{sec:finite-temp}

\subsection{Simulation parameters}
The simulation parameters ($\beta$ and $N_\tau$) for several $T/T_c$ are determined by using Eq.(\ref {eq:t0-beta}) as shown in~Table~\ref{table:param-finiteT}.
The value of $T_c=[N_\tau a]^{-1}$ is fixed at~$\beta=2.4265, N_\tau=6$, which is given in~Ref.~\cite{Engels:1992fs}.
The estimated values of $\beta$ at $T_c$ on the other $N_\tau$ are obtained by Eq.(\ref {eq:t0-beta}) as $\beta_c=2.514, 2.583,$ and $2.650$ for $N_\tau=8,10,$ and $12$, respectively.
On the other hand, the direct measurement of the critical values of $\beta$, which is determined by the Polyakov-loop susceptibility for each $N_\tau$, is summarized in Ref.~\cite{Berg:2016wfw}; $\beta_c=2.510363(71), 2.57826(14)$, and $2.63625(35)$ for $N_\tau=8,10,$ and $12$, respectively.
Our estimated values are consistent with the direct determinations in the two- or three-digit accuracy.
In this work, we determine the value of $\beta$ for each temperature in the three-digit accuracy in the finite-temperature simulation.

\begin{table}[h]
\begin{center}
\begin{tabular}{|c||c|c|c|c|}
\hline
 $T/T_c$  &  $N_\tau=6$  &    $N_\tau=8$  &    $N_\tau=10$  &  $N_\tau=12$ \\
 \hline
 0.95  &   (2.41)   &   2.50  &     2.57 &  2.62 \\
 0.98  &   2.42   &   2.51  &     2.58 &  2.63 \\
 1.01  &   2.43   &   2.52  &     2.59 &  2.64 \\
 1.04  &   2.44   &   2.53  &     2.60 &  2.65 \\
 1.08  &   2.45   &   2.54  &     2.61 &  2.66 \\
 1.12  &   2.46   &   2.55  &     2.62 &  2.67 \\
 1.28  &   2.50   &   2.59  &     2.66 &  2.72 \\
 1.50  &   2.55   &   2.64  &     2.71 &  2.77 \\
 1.76  &   2.60   &   2.69  &     2.76 &  2.82 \\
 2.07  &   2.65   &   2.74  &     2.81 &  (2.87) \\
\hline
\end{tabular}
\caption{Lattice parameters ($\beta$ for each $N_\tau$) at finite temperature simulations.  $\beta=2.41$ and $2.87$ in the bracket are outside of the interpolating regime of the scale-setting function (Eq.~(\ref {eq:t0-beta})). We did not use these parameters in the main analyses. }  \label{table:param-finiteT}
\end{center}
\end{table}
The thermodynamic quantities have been obtained using $200$ configuration with $100$-sweep separations for each lattice parameter.
The interval of the gradient flow-time for the measurement of the EMT is $\Delta t/a^2=0.01$.

\subsection{Simulation results: Thermodynamic quantities}\label{sec:thermo}
The procedure to calculate the EMT on the lattice is summarized as following four steps in Ref.~\cite{Asakawa:2013laa}; \\
\textbf{Step 1:} Generate gauge configurations at~$t=0$ on a space-time lattice
with the lattice spacing~$a$ and the lattice size~$N_s^3\times N_\tau$.\\
\textbf{Step 2:} Solve the gradient flow for each configuration to obtain the
flowed link variables in the fiducial window, $a\ll\sqrt{8t}\ll R$, to suppress the discretization and the finite volume effects. Here,
$R$~is an infrared cutoff scale such as~$\Lambda_{\text{QCD}}^{-1}$
or~$T^{-1}=N_\tau a$. \\
\textbf{Step 3:} Construct $U_{\mu\nu}(t,x)$ and~$E(t,x)$
in~Eq.~(\ref{eq:def-U-E}) using the flowed link variables
and average them over the gauge configurations at each~$t$.\\
\textbf{Step 4:} 
Carry out an extrapolation toward $(a,t)=(0,0)$, first $a\to0$ and then $t\to0$ under the condition in \textbf{Step 2}. 

We have to carefully estimate the propagation of errors, in particular, taking the double-limits in \textbf{Step 4}, since each flow-time data after taking the continuum extrapolation is correlated with each other.
In this work, we use the jackknife method. Thus, firstly we generate the jackknife samples for the lattice raw data of $U_{\mu \nu}$ and $E$ in each lattice parameter, and independently obtain the renormalized EMT by taking the double limits for each jackknife sample. 
Finally, we calculate the standard error from the deviation of the obtained results.

\begin{figure}[h]
\begin{center}
 \includegraphics[width=0.9\textwidth,clip]{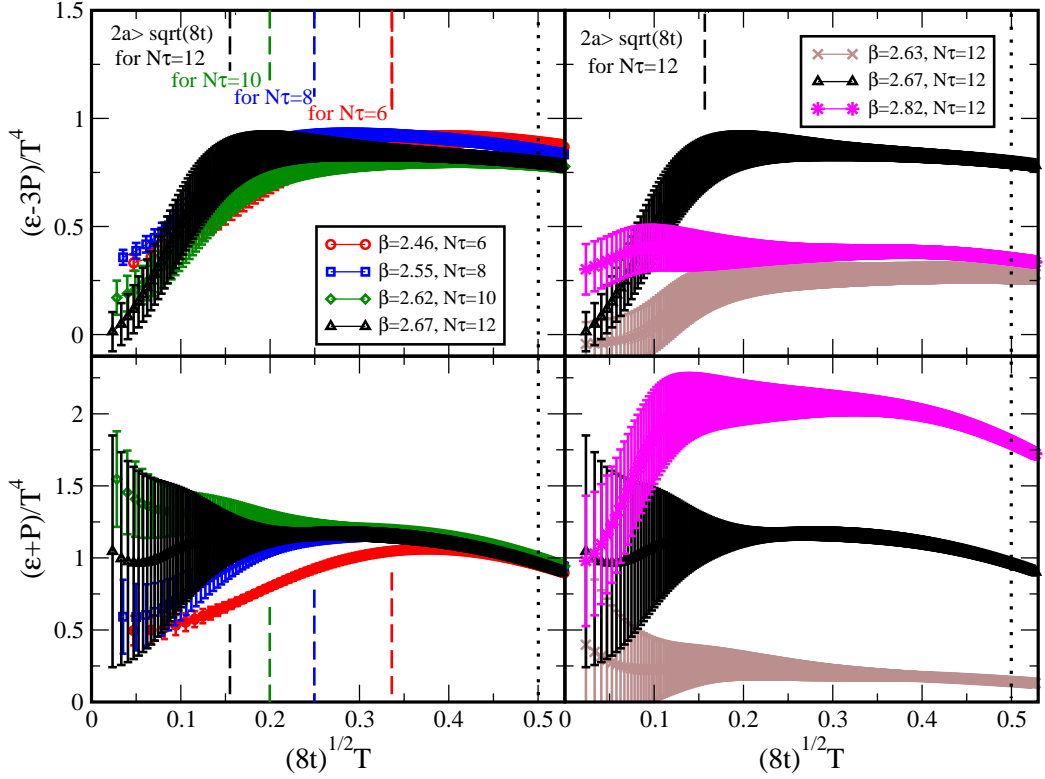}
\caption{Flow-time dependence for the expectation values of the trace anomaly (top two panels) and entropy density (bottom two panels) in each lattice parameter. Left two panels show the $N_\tau$ (lattice spacing) dependence at $T=1.12T_c$. Right two panels show the temperature dependence at the fixed $N_\tau=12$. }
\label{fig:raw-data}
\end{center}
\end{figure}
The left two panels in Fig.~\ref{fig:raw-data} show the dimensionless trace anomaly ($\Delta/T^4=(\varepsilon-3P)/T^4$) and the
dimensionless entropy density ($s/T^3=(\varepsilon+P)/T^4$) at~$T=1.12T_c$ as a function of the dimensionless flow parameter~$\sqrt{8t}T$.
Here, the error bar denotes the statistical errors.
The lower limit of the fiducial window in~\textbf{Step 2} is indicated by the dashed
lines in~Fig.~\ref{fig:raw-data}. This lower limit, which is related to the discretization error, is set to
be~$\sqrt{8t_{\rm min}}=2a$, since we consider the clover-leaf operator with a size~$2a$. 
The upper limit of the fiducial window is shown in the dotted line.
It is set to be $\sqrt{8t_{\rm max}}=0.35N_\tau a$~\cite{Eller:2018yje}, so that the smearing by the gradient flow does not exceed the temporal lattice size.
The data with statistical errors in~Fig.~\ref{fig:raw-data} show the plateau inside the fiducial window
($2/N_\tau \le \sqrt{8t}T \le 0.35$) for each~$N_\tau$.
The right two panels in~Fig.~\ref{fig:raw-data} present the temperature dependence of raw data in $N_\tau=12$, and we can find that the plateaus inside the fiducial window are shown for all temperature regimes.
It suggests that the contributions of the higher dimensional operator, which is proportional to $t$, are small in Eq.~(\ref{eq:E-expand-t}).

\begin{figure}[h]
\begin{center}
\includegraphics[width=0.9\textwidth,clip]{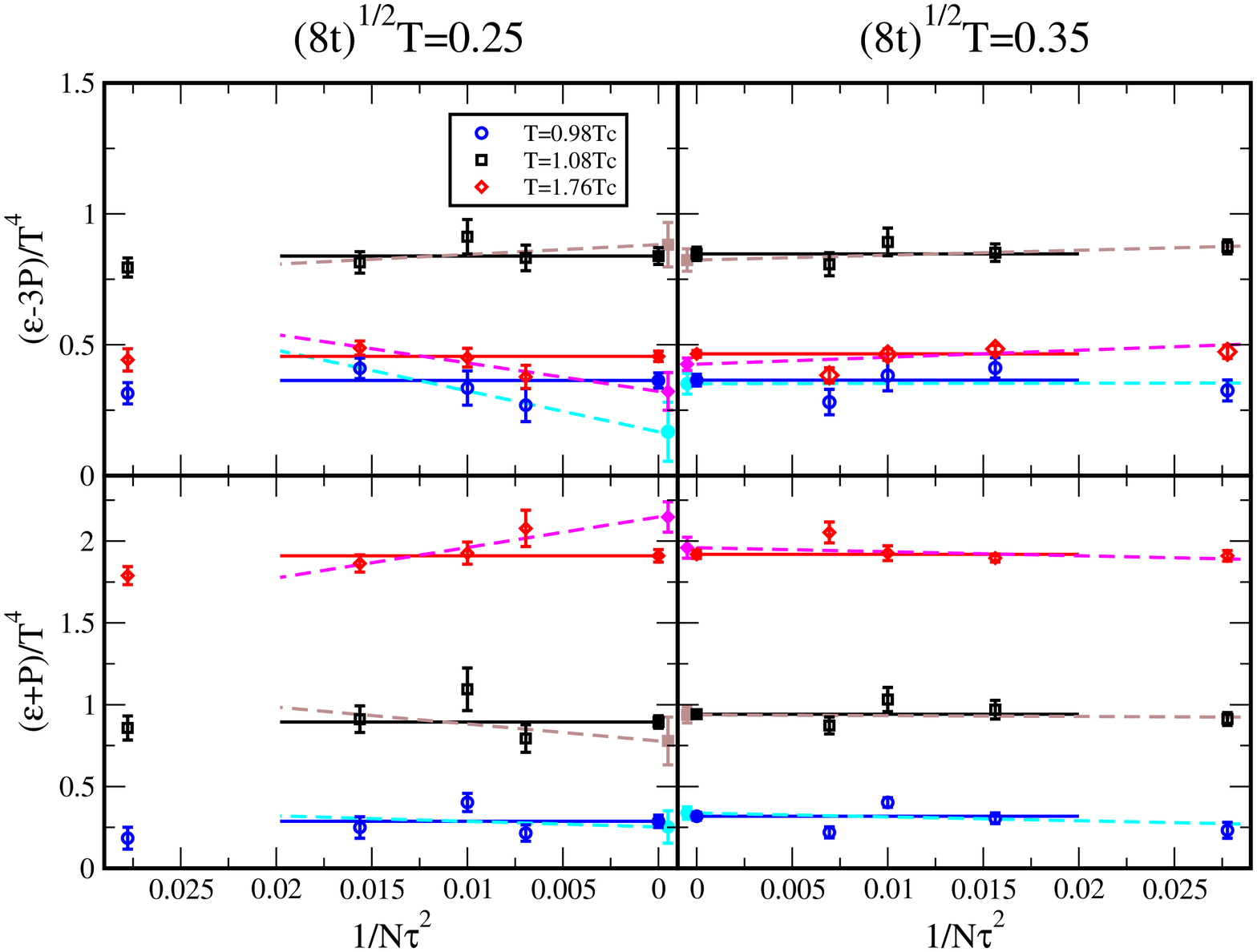}
\caption{Continuum extrapolation for the fixed flow-time $\sqrt{8t}T=0.25$ and $0.35$. The solid and dashed lines denote the constant and linear extrapolations of $(Ta)^2=1/{N_\tau}^2$, respectively. Note that the continuum limit directs to the center-line of the figure. To estimate the systematic error, the four data points are used at $\sqrt{8t}T=0.35$, while only the three data are fitted at $\sqrt{8t}T=0.25$, since the data of $N_\tau=6$ are outside of the fiducial window. }
\label{fig:cont-lim}
\end{center}
\end{figure}
In \textbf{Step 2}, we carry out both constant and linear extrapolations for $160$ datasets, namely $16$ fixed flow-time in increments of $0.01$ from $0.25$ to $0.35$ for each temperature listed in Table~\ref{table:param-finiteT}.
Figure~\ref{fig:cont-lim} shows the example plots of these continuum extrapolations at the fixed flow-time for $T=0.98T_c, 1.08T_c,$ and $1.76T_c$.
The solid and dashed lines denote the constant and linear extrapolations of $(Ta)^2=1/{N_\tau }^2$, respectively.
As central analyses, we utilize the constant extrapolation of three data, $N_\tau=8,10$, and $12$, for $0.95 \le T/T_c \le 1.76$. 
Only for $T = 2.07T_c$, where the number of the data points is few and we expect that the discretization error in $N_\tau = 6$ would be acceptable, the constant extrapolations by two data ($N_\tau = 8$ and $10$) in the short flow-time regime ($\sqrt{8t}T<1/3$) and by three data ($N_\tau = 6,8$, and $10$) in the long flow-time regime ($\sqrt{8t}T \ge 1/3$), are carried as central analyses.
The linear-fit results are used to estimate the systematic error from the scaling violation.
In the continuum limit, two extrapolated values of the trace anomaly (the entropy density) are consistent each other within $2$--$\sigma$ ($3$--$\sigma$), where $\sigma$ denotes the statistical standard error.

\begin{figure}[h]
\begin{center}
\includegraphics[width=0.9\textwidth,clip]{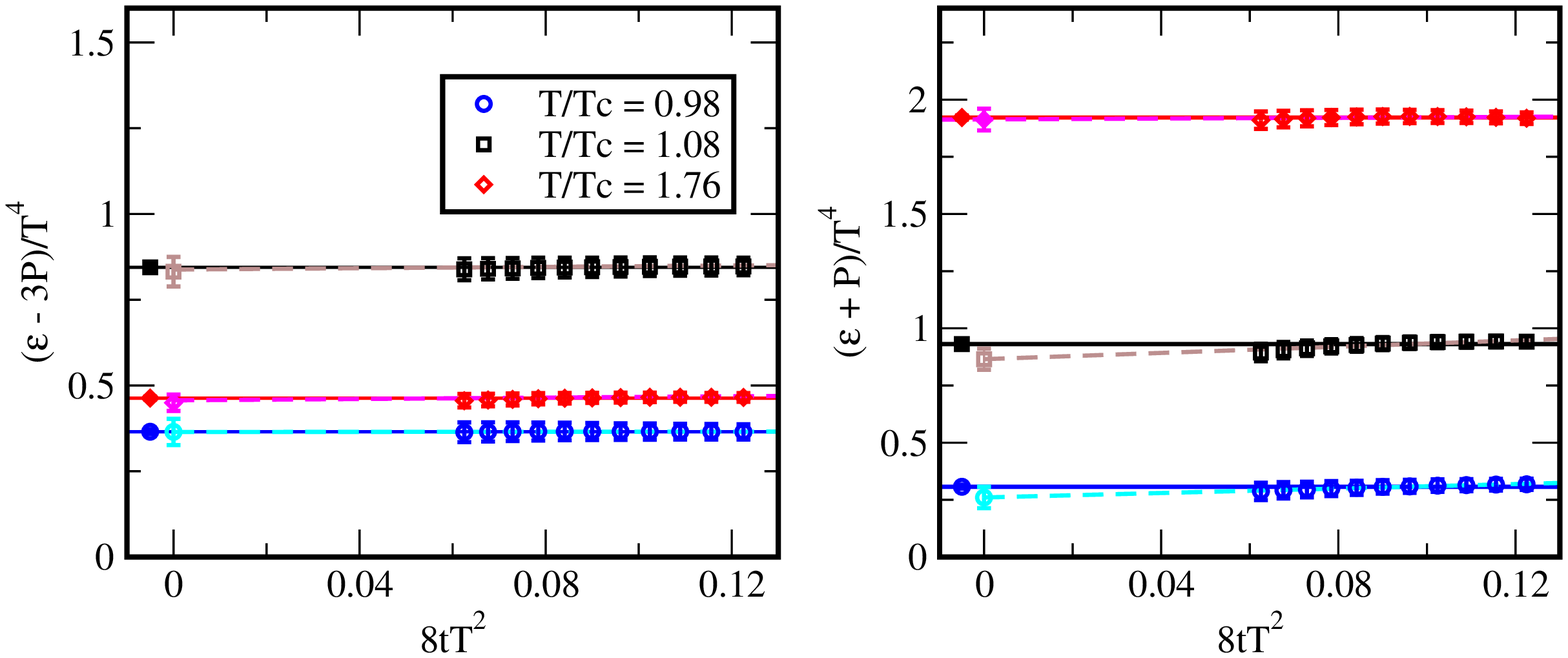}
\caption{Plots of the extrapolations in the $t \rightarrow 0$ limit. The flow-time ($\sqrt{8t}T$) takes from $0.25$ to $0.40$ in increments of $0.01$.
Each data point is obtained after taking the continuum extrapolation, which is independently carried out at each fixed flow-time. The solid and dashed lines denote the constant and linear extrapolations in terms of $t$, respectively. }
\label{fig:cont-t-zero-lim}
\end{center}
\end{figure}
Figure~\ref{fig:cont-t-zero-lim} shows the $t \rightarrow 0$ extrapolation.
The solid and dashed lines denote the constant and linear extrapolations of $t$, respectively.
We take the result, which has the better reduced $\chi^2$, as central values, and estimate a systematic uncertainty from the discrepancy depending on the extrapolation function. 
The systematic error for the trace anomaly is almost the same as the statistical error, while that of the entropy density is at most three times as large as the statistical one.

We also estimate the other systematic error coming from the uncertainty of the lattice determination of $\Lambda_{\overline{\text{MS}}}$.
It changes the values of the coefficients $\alpha_U(t)$ and $\alpha_E(t)$ via the renormalized coupling constant, and the difference becomes larger in the lower temperature regime since the running coupling constant rapidly grows up in that regime.
The corresponding systematic error for the trace anomaly is negligible in comparison with the statistical error, while the one for the entropy density changes the central value at most by $6\%$.

\begin{figure}[h]
\begin{center}
\includegraphics[width=0.9\textwidth,clip]{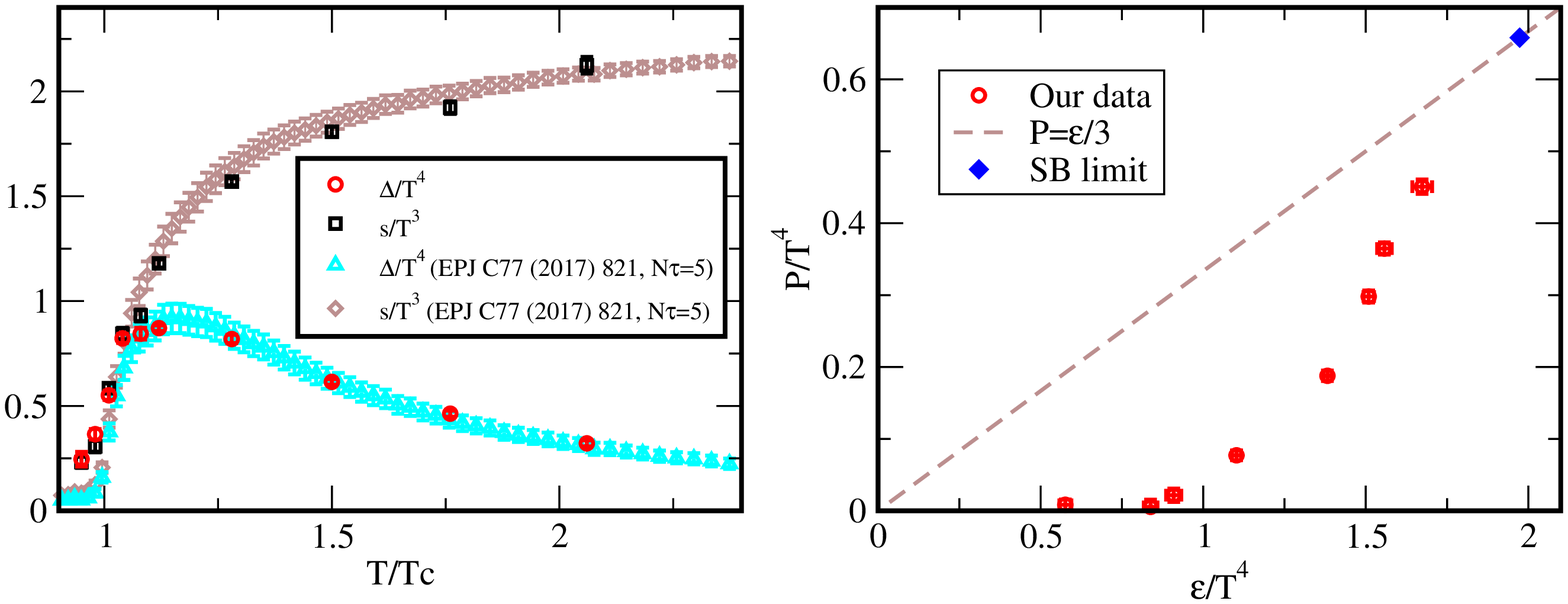}
 \caption{(Left)~Results of the trace anomaly (circle-red) and the entropy density (square-black) as a function of temperature.
 The triangle (cyan) and diamond (brown) symbols denote the trace anomaly and the entropy density at $N_\tau = 5$ given in Ref.~\cite{Giudice:2017dor}, respectively. All error bars denote only the statistical errors.
 (Right)~The equation of state, namely the relationship between the energy density and the pressure. The dashed line, $P=\varepsilon/3$, denotes the case with vanishing trace anomaly. The diamond (blue) symbol represents the SB limit $(\varepsilon/T^4,P/T^4)=(\pi^2/5,\pi^2/15)$.}
\label{fig:traceanomaly-entropy}
\end{center}
\end{figure}
We plot the trace anomaly (red-circle symbols) and entropy density (black-square symbols) as a function of $T/T_c$ after taking the double $(a,t) \rightarrow (0,0)$ limits  in the left panel of Fig.~\ref{fig:traceanomaly-entropy}.
We also summarize the data in Appendix~\ref{sec:data}.
Here, the error bars in the figure and the table denote only statistical errors.
As a comparison, the trace anomaly (cyan-triangle symbols) and the entropy density (brown-diamond symbols), which are obtained by the integral method using the improved action~\cite{Giudice:2017dor} at the fixed temporal lattice extent $N_\tau=5$, are also shown in the figure.
In $T/T_c \ge 1.12$, our results including the above three types of systematic errors agree with the results given by the integral method. 
Note that our statistical errors obtained by a fewer number of configurations are smaller than the ones in the integration methods.
In low temperature regime, there are small discrepancies among ours and the results given in Refs.~\cite{Caselle:2015tza, Giudice:2017dor}.
Our data of the trace anomaly are larger than the results in Ref.~\cite{Caselle:2015tza}, which utilizes the same lattice action and takes the continuum extrapolation.
Furthermore, the result in Ref.~\cite{Caselle:2015tza} is larger than the one in Ref.~\cite{Giudice:2017dor}, which utilizes the improved action and the fixed $N_\tau$.
There are several possible reasons for the discrepancy, which appears especially in the low temperature.
The one would come from the discretization errors, that exist in both methods, and it may be needed to take the extrapolations more carefully.
The other reason for the gradient flow method would come from the usage of the one-loop approximation.

The right panel of Fig.~\ref{fig:traceanomaly-entropy} shows the equation of state in $T \ge T_c$, namely the relationship between the energy density and the pressure. 
The linear function, $P=\varepsilon /3$, presents the case with vanishing trace anomaly, and the diamond (blue) symbol denotes the values in the ideal gas (Stefan-Boltzmann (SB)) limit.
The nonperturbative interaction reduces the values of pressure in the whole energy-density regime.  
In high temperature regime, our result heads toward the point in the SB limit, but still, the lattice data with the highest $\varepsilon$, which correspond to $T \simeq 2T_c$, are almost $70$--$85\%$ of the value in the limit.
It is evidence that the state of the two-color ``QGP'' phase around $T \le 2T_c$ cannot be described by the ideal gas model yet.

\begin{figure}[h]
\begin{center}
\includegraphics[width=0.9\textwidth,clip]{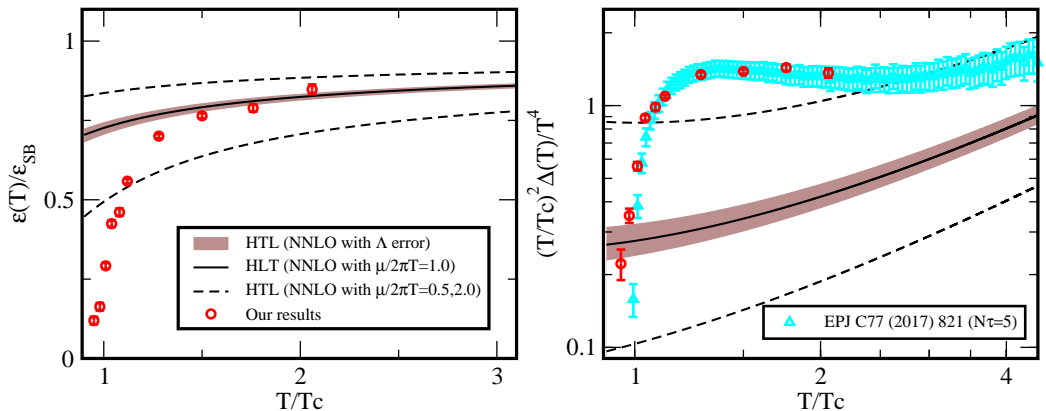}
 \caption{(Left) Results of the energy density normalized by those in the SB limit. (Right) Rescaled trace anomaly, $(T/T_c)^2\Delta(T)/T^4$. 
 As a comparison, the Hard-Thermal-Loop (HTL) results in the next-next-to-leading order (NNLO) are also shown~\cite{Andersen:2010ct}. 
 In both panels, the solid (black) curve denotes the result of HTL with $\mu_{HTL}/(2 \pi T)=1.0$, and the two dashed curves present the ones with $\mu_{HTL}/(2 \pi T)=0.5$ and $2.0$, respectively. The brown bounds show the HTL results with the error bar coming from $T_c/\Lambda_{\overline{\text{MS}}}=1.23(11)$.
The triangle (cyan) symbols in the right panel denote the results of the trace anomaly at $N_\tau = 5$ given in Ref.~\cite{Giudice:2017dor}.}
\label{fig:energy-density}
\end{center}
\end{figure}
Now, let us compare our results with the analytic prediction, namely the results of the Hard-Thermal-Loop (HTL) model. 
The left panel of Fig.~\ref{fig:energy-density} shows the comparison of the energy density normalized by the value in the SB limit between our lattice result and the HTL calculation in $N_c=2$ case~\cite{Andersen:2010ct}.
In HTL analyses, we use the next-next-to-leading order (NNLO) formula with the three-loop running coupling constant shown in Eq.~(\ref{eq:alpha-s}) with the renormalization scheme $\mu_{HTL}=2\pi T$.
In the calculation, we use $T_c/\Lambda_{\overline{\mathrm{MS}}}=1.23(11)$ obtained by the lattice data.
The (brown) bound in Fig.~\ref{fig:energy-density} shows the uncertainty coming from the error in $T_c/\Lambda_{\overline{\mathrm{MS}}}$, where the renormalization scale is fixed as $\mu_{HTL}/(2\pi T)=1.0$.
The systematic uncertainty of the choice of $\mu_{HTL}$ is shown as two dashed curves obtained by using $\mu_{HTL}/(2 \pi T)=0.5$, and $2.0$.
Our results are nicely consistent with the HTL results until near $T_c$. 

Finally, to see the scaling law of the trace anomaly more precisely, we also compare the results between the lattice data and the HTL analyses.
The trace anomaly has a leading correction term of $1/T^2$ in the high temperature regime, as we have shown in Fig.~\ref{fig:anomaly-vs-Panero}.
The data can be well-fitted by the linear function of $1/T^2$ in $(T_c/T)^2 \le 0.6$ regime.
On the other hand, the nonperturbative logarithmic correction term for $\Delta (T)/T^2$ is predicted by the HTL analyses.
To see these correction terms, we take both axes to a logarithmic scale in the right panel of Fig.~\ref{fig:energy-density}.
In $1.3T_c \lesssim T$, the lattice data exhibit almost plateau and approaches to the HTL results.  
In the SU($3$) gauge theory~\cite{Borsanyi:2012ve}, the similar behavior also appears around $T = 2T_c$, and in the further high temperature the lattice data become consistent with the HTL and the perturbative analyses.
We may consider that it also occurs in the SU($2$) gauge theory.

\section{Summary and future directions}\label{sec:summary}
In this work, we numerically investigate the thermodynamics of the pure SU($2$) gauge theory.
The theory is a good testing ground for the studies on the methodology and the $N_c$-dependence for the pure SU($N_c$) and QCD theory.
We determine the scale-setting function and thermodynamic quantities by using the gradient flow method in the $N_c=2$ case.

For the precise scale-setting of lattice parameters, we propose a reference scale $t_0$ for the SU($2$) gauge theory, which satisfies $t^2 \langle E \rangle |_{t=t_0} = 0.1$. 
This value is determined by a natural scaling-down of the standard $t_0$-scale for the SU($3$) gauge theory.
We have shown that the reference scale is suitable to study thermodynamics in the temperature region of $T_c\lesssim T \lesssim 2T_c$.

We also obtain the thermodynamic quantities, which are directly calculated from the EMT by the small flow-time expansion of the gradient flow method.
This work is the first application of the gradient flow method to the thermodynamic quantities for the SU($2$) theory.
We take the double limits ($a, t$) $\rightarrow$ ($0,0$), first $a\rightarrow 0$ and then $t \rightarrow 0$, to remove the artifact in the gradient flow method. 
The statistical error is smaller than the one given by the integration method, nevertheless, we utilize only a few hundred configurations.
We precisely show that the trace anomaly in the pure SU($2$) gauge theory has a different scaling property from the $N_c \ge 3$ cases.
It shows the gently curved behavior near $T_c$, and the linear behavior of ($1/T^2$) appears in $(T_c/T)^2 \le 0.6$ regimes.
We also find a strong tendency toward the consistency with the HTL results in the high-temperature regime. \\

For future works, we address the following points.\\
\noindent{\bf Universality of the critical phenomena}\\
The finite-temperature phase transition in the pure SU($2$) gauge theory is characterized by the Polyakov loop, and it is $Z_2$-symmetric/ broken phase transition.
It is believed that the phase transition belongs to the same universality class as the Ising model in three dimensions since it has the same symmetry and spatial dimensions~\cite{Svetitsky:1982gs}.
However, it is hard tasks to obtain precisely the same critical exponents with the one for the Ising model.
Actually, we also observed the critical exponent of the scaling of the Polyakov loop, but still it is the similar value with the one in Ref.~\cite{Hubner:2008ef} and is not consistent with the value of the Ising model, $\beta=0.3265(3)$.
On the other hand, as for trace anomaly, 
the recent lattice Monte Carlo study on the $3$d Ising model shows the evidence of the traceless of the EMT~\cite{Meneses:2018xpu}. 
Meanwhile, the trace anomaly of the pure SU($2$) gauge theory at the critical temperature with one compact-dimension has a finite value. 
However, this is not so strange, since the EMT can be affected by the all scale fluctuations near the transition point and is not expected
to behave universally as in the $3$d Ising model. 

\noindent{\bf Exact determination of $\alpha_U, \alpha_E$ coefficients}\\
In our analysis, we utilize the one-loop calculation for the coefficients; $\alpha_U$ and $\alpha_E$.
Our results are consistent with the ones given by the integral methods in the high temperature regime, but there is a small discrepancy in $T< T_c$ even if the systematic errors are included.  
The discrepancy may come from the usage of the one-loop approximation in these coefficients. 
The analytical calculation of these coefficients in the higher order is future work. 
Furthermore, the nonperturbative determination of these coefficients has been proposed~\cite{DelDebbio:2013zaa, Capponi:2016yjz}. That is also a valuable direction.

\noindent{\bf Calculation of viscosities}\\
One of the motivations for this work is to prepare the determination of $\eta/s$, where $\eta$ is the shear viscosity.
The SU($N_c$) gauge theory with large-$N_c$ has a lower bound of $\eta/s=1/(4\pi)$ based on the AdS/CFT correspondence.
It is also suggested that the $\eta/s$ takes the minimum value at the phase transition in a wide class of the systems~\cite{Chen:2010vf}.
However, the $1/N_c$ corrections, even their signs of the correction terms, are unclear~\cite{Kats:2007mq}.
The lattice results on the pure SU($3$) approach the bound near the critical temperature~\cite{Mages:2015rea, Astrakhantsev:2017nrs}, and $\eta/s$ on the pure SU($2$) at $T\simeq 1.2 T_c$ is also very close to the bound~\cite{Astrakhantsev:2015jta}.
The systematic study on the temperature-dependence and the $N_c$-dependence could reveal properties of the vacuum in the QCD(-like) theories via these viscosities.

\noindent{\bf Constructing an effective model of two color QCD }\\
Once the temperature dependences of the
energy density (or pressure) and the expectation value of the Polyakov loop are determined by the pure gauge lattice simulations, 
one can construct the effective Polyakov-loop potential used in effective model such as the PNJL model~\cite{Meisinger:1995ih,Dumitru:2002,Fukushima:2003fw,Brauner:2009gu,Ratti:2005jh,Megias:2004hj}.  
Using the effective model, we can analyze the physics in full QCD which contains dynamical quark contributions. 
Since two-color full QCD simulation is easier to be treated than three color one, it is also easier to check the efficiency of the constructed effective model in two color case. 
In particular, it is able to investigate the efficiency even at finite quark number density~\cite{Kashiwa:2012xm}, since the lattice simulation of two-color full QCD is also feasible at finite quark chemical potential.

\section*{Acknowledgments}
We would like to thank M. Yahiro and M.~Yamazaki deeply for valuable comments and discussions.
E.I. would like to thank K.~Iida and M.~Panero for useful comments. 
H.K would like to thank H.~Yoneyama for valuable discussions. 
Numerical simulations were performed on xc40 at YITP, Kyoto University and on SX-ACE at the Research Center for Nuclear Physics (RCNP), Osaka University.
The work of H.~K. is supported in part by a Grant-in-Aid for Scientific Researches No.~17K05446. 

\appendix
\section{Distribution of the topological charge at finer lattice}\label{sec:topology}
To investigate the autocorrelation of the generated configuration, we measure the topological charge using the gradient flow.
The topological charge is related with the vacuum structure, and it has a long autocorrelation among the observables in Yang-Mills theory. 
The gluonic definition of the topological charge in Euclidean space-time is given by
\beq
Q=  \frac{1}{32\pi^2 } \int d^4x \epsilon_{\mu \nu \rho \sigma} \mathrm{Tr} G^a_{\mu \nu} G^a_{\rho \sigma},\label{eq:def-topology}
\eeq
where $\epsilon_{\mu \nu \rho \sigma}$ denotes the totally antisymmetric tensor.
In this gluonic definition, the topological charge of quantum configurations in lattice simulations generally does not take an integer-value because of UV fluctuations. 
The application of the gradient flow suppresses the UV fluctuations and recovers almost integer-valued quantity~\cite{Luscher:2010iy,Bonati:2014tqa}.

\begin{figure}[h]
\begin{center}
\includegraphics[scale=0.4,clip]{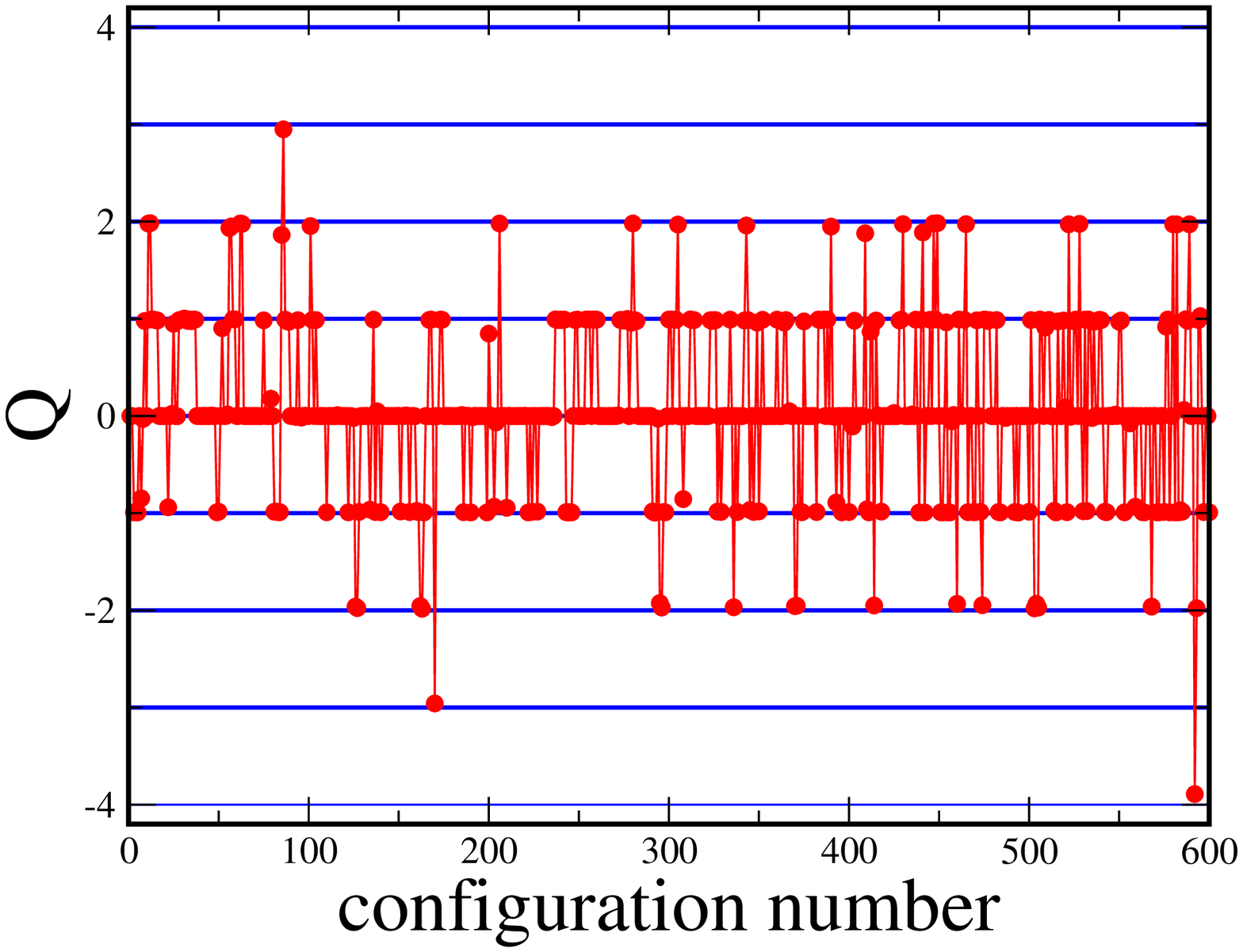}
\caption{History of topological charge at $\beta=2.85$ obtained by the gradient flow method. Horizontal axis denotes configuration number, and each configuration separates $100$ sweeps.}
\label{fig:topology}
\end{center}
\end{figure}
Figure~\ref{fig:topology} shows the topological charge for each $600$ configuration in $\beta=2.85$, in which the lattice spacing is the smallest and being most likely to occur a topological freezing in our simulations. 
Here, we observe the topological charge at the gradient flow-time $t/a^2=32$, where the effective smeared-regime is the same with the half of the lattice extent.
We found that the topological charge takes an almost integer-value as expected for each configuration and it frequently changes within one configuration separation.
We conclude that the autocorrelation can be negligible in our data sets. 

\section{Scale-setting function for several reference scales}\label{sec:other_t0_scale}
In our analysis in \S.4.2, the scale-setting function is valid for $2.420 \le \beta \le 2.850$.
The range is determined by the choice of the reference scale ($t_0$) to avoid a strong lattice artifact ($t_0/a^2 \ge 1$) and a finite-volume effect ($\sqrt{8 t_0/a^2} \le N_\tau /2$) with the reference value $A \equiv t^2\langle E \rangle|_{t=t_0} =0.1$.
However, if we take the other reference values, then we can extend the available regime of $\beta$.

\begin{table}[h]
 \centering
 \begin{tabular}[t]{|c||c|c|c|c|}
  \hline
  & \multicolumn{4}{c|}{$t_0/a^2$} \\ \hline 
  $\beta$ & $A=0.08$& $A=0.09$  & $A=0.10$  & $A=0.12$ \\ \hline
  2.400 & 0.7264(9)  & 0.8400(12)& 0.9549(5) & 1.171(23) \\
  2.420 & 0.8202(11) & 0.9529(14)& 1.083(2)  & 1.336(2) \\
  2.500 & 1.372(2)   & 1.609(2)  & 1.839(3)  & 2.279(4) \\
  2.600 & 2.612(7)   & 3.075(8)  & 3.522(10) & 4.370(14)\\
  2.700 & 4.881(25)  & 5.770(33) & 6.628(36) & 8.247(54)\\
  2.800 & 8.780(74)  & 10.40(10) & 11.96(12) & 14.92(17)\\
  2.850 & 12.25(10)  & 14.63(14) & 16.95(17) & 21.43(25)\\ 
  \hline
 \end{tabular}
 \caption{Results of $t_0/a^2$ for $A = 0.08$, $0.09$, $0.10$ and $0.12$.}
 \label{tab:other_t0_scale}
\end{table}
Table~\ref{tab:other_t0_scale} shows the values of $t_0$ in lattice units for $A = 0.08$, $0.09$, $0.10$ and $0.12$.
In $\beta=2.400$ and $2.420$, the proper reference values should be taken in $A \ge 0.12$ and $A \ge 0.10$, respectively.

To see the scale violation coming from the choice of the reference values, we show the ratio of the lattice spacing ($a/a_0$) for each $\beta$ in Table~\ref{tab:other_t0_scale_2}.
Here, we rescale the lattice spacing with the value at $\beta = 2.600$ for each $A$. 
 \begin{table}[h]
  \centering
  \begin{tabular}{|c||c|c|c|c|}
   \hline
   & \multicolumn{4}{c|}{$a/a_0$} \\ \hline
   $\beta$& $A=0.08$& $A=0.09$     & $A=0.10$   & $A=0.12$  \\ \hline
   2.400 & 1.896(3)  & 1.913(3)     & 1.921(3)   & 1,932(4)  \\
   2.420 & 1.784(3)  & 1.797(3)     & 1.803(3)   & 1.809(3)  \\
   2.500 & 1.380(2)  & 1.383(2)     & 1.384(2)   & 1.385(3)  \\
   2.600 & 1.000(2)  & 1.000(2)     & 1.000(2)   & 1.000(2)  \\
   2.700 & 0.7315(21)& 0.7301(23)   & 0.7290(22) & 0.7279(27)\\
   2.800 & 0.5454(24)& 0.5438(26)   & 0.5427(28) & 0.5412(31)\\
   2.850 & 0.4617(20)& 0.4586(22)   & 0.4558(24) & 0.4516(18)
   \\ \hline
  \end{tabular}
  \caption{Ratios of the lattice spacings $a/a_0$ for $A = 0.08$, $0.09$, $0.10$ and $0.12$, where we take the reference lattice spacing $a_0$ with the value at $\beta=2.600$.}
  \label{tab:other_t0_scale_2}
 \end{table}
Table ~\ref{tab:other_t0_scale_2} indicates that the relation between $\beta$ and $a$ is almost independent of the choice of the reference values.

Finally, we obtain the scale-setting function with the following interpolating function;
\begin{eqnarray}
 \label{eq:other_t0_scale_1}
  \ln(t_0/a^2) = \alpha_0 + \alpha_1(\beta - 2.600) + \alpha_2(\beta - 2.600)^2,
\end{eqnarray}
where $\alpha_0$, $\alpha_1$ and $\alpha_2$ are the fitting parameters.
The obtained parameters and the fit range for each reference value are summarized in Table~\ref{tab:other_t0_scale_3}.
Although the constant term $\alpha_0$, which corresponds to $\ln (t_0/a^2)$ at $\beta=2.600$, depends on the choice of $A$, the other coefficients $\alpha_1$ and $\alpha_2$ do not change so much. 
It suggests that we can connect several scale-setting functions obtained by the different choice of $A$, and can obtain the extended scale-setting function, which is available in the wider regime. 
 \begin{table}[h]
  \centering
  \begin{tabular}{|c||c|c|c|c|}
   \hline
   $A$  & $\alpha_0$   & $\alpha_1$& $\alpha_2$    & fit range  \\ \hline
   0.08   & 0.9590(22) & 6.343(16) & -0.8548(1360) & [2.500,2.850]\\
   0.09   & 1.122(2)   & 6.382(18) & -0.8271(1493) & [2.500,2.850]\\
   0.10   & 1.258(2)   & 6.409(14) & -0.7411(851)  & [2.420,2.850]\\
   0.12   & 1.474(2)   & 6.439(15) & -0.7501(846)  & [2.400,2.850]\\ \hline
  \end{tabular}
  \caption{The coefficients of the scale-setting function for $A = 0.08$, $0.09$, $0.10$ and $0.12$.}
  \label{tab:other_t0_scale_3}
 \end{table}

\section{Data}\label{sec:data}

\begin{table}[h]
\begin{center}
\begin{tabular}{|c|c|c|c|c|}
\hline
$T/T_c$  &    $\Delta/T^4$ &  $s/T^3$ &  $\varepsilon/T^4$ & $P/T^4$ \\ \hline
0.95  &  0.246(35) & 0.232(27) & 0.235(24) & -0.00348(1041) \\
0.98  &  0.365(25) & 0.307(29) & 0.322(24) & -0.0145(87)    \\
1.01  &  0.551(22) & 0.585(22) & 0.576(18) & 0.00848(739)   \\
1.04  &  0.822(22) & 0.844(29) & 0.838(21) & 0.00543(1019)  \\
1.08  &  0.844(28) & 0.931(28) & 0.909(23) & 0.0217(93)     \\
1.12  &  0.871(13) & 1.18(2)   & 1.10(1)   & 0.772(69)      \\
1.28  &  0.819(20) & 1.57(2)   & 1.38(1)   & 0.188(7)       \\
1.50  &  0.615(13) & 1.81(2)   & 1.51(2)   & 0.298(8)       \\
1.76  &  0.463(14) & 1.92(3)   & 1.56(2)   & 0.365(9)       \\
2.07  &  0.322(14) & 2.12(4)   & 1.67(3)   & 0.451(10)      \\
\hline
\end{tabular}
\caption{Data of thermodynamic quantities; trace anomaly ($\Delta/T^4$), entropy density ($s/T^3$), energy density ($\varepsilon/T^4$), pressure ($P/T^4$).
The errors are statistical ones.
} \label{table:thermo-data}
\end{center}
\end{table}

\end{document}